\documentclass[12pt, prd,onecolumn,floatfix,letterpaper,amsmath,amssymb,nofootinbib,superscriptaddress]{revtex4}
\usepackage{graphicx}
\usepackage{amsmath}
\usepackage{amssymb}
\usepackage[footnotesize]{caption}

\begin{document}

   \title{Power-counting and Renormalizability in Lifshitz Scalar Theory}

   \author{Toshiaki Fujimori}
   	\email{toshiaki.fujimori018@gmail.com}
    \address{Department of Physics, National Taiwan University, Taipei 10617, Taiwan, R.O.C.}
 
 \author{Takeo Inami}
   	\email{inami@phys.chuo-u.ac.jp}
 \address{Department of Physics, National Taiwan University, Taipei 10617, Taiwan, R.O.C.}
\address{Mathematical Physics Lab., Riken Nishina Center, Saitama 351-0198, Japan}

   \author{Keisuke Izumi}
   	\email{izumi@phys.ntu.edu.tw}
    \address{Leung Center for Cosmology and Particle Astrophysics, National Taiwan University, Taipei 10617, Taiwan, R.O.C.}
  
     \author{Tomotaka Kitamura}
   	\email{kitamura@gravity.phys.waseda.ac.jp}
    \address{Department of physics, Waseda University, Shinjyuku, Tokyo 169-8555, Japan}   


\begin{abstract}
\vspace{5mm}
We study the renormalizability in theories of a self-interacting Lifshitz scalar field. 
We show that although the statement of power-counting is true at one-loop order, 
in generic cases where the scalar field is dimensionless, 
an infinite number of counter terms are involved in the renormalization procedure. 
This problem can be avoided by imposing symmetries, the shift symmetry in the present paper, 
which allow only a finite number of counter terms to appear. 
The symmetry requirements might have important implications 
for the construction of matter field sectors in the Ho{\v r}ava-Lifshitz gravity.
\end{abstract}

\maketitle

\section{Introduction}
In 1941, Lifshitz proposed a scalar field theory with an unconventional scaling property
to describe tri-critical phenomena in condensed matter physics \cite{Lifshitz scalar}. 
The scalar theory, called the Lifshitz scalar theory, has an improved UV behavior 
due to the following anisotropic scaling of space and time  
\begin{eqnarray}
t \rightarrow b^z t,  \hspace{2cm} x^i \rightarrow bx^i \hspace{5mm} (i=1,\dots d). \label{1}
\end{eqnarray}
This scaling is called the Lifshitz scaling. 
The integer $z$ denotes the dynamical critical exponent, 
which indicates the degree of anisotropy between space and time. 
Because of this anisotropic scaling, the Lorentz symmetry is explicitly broken for $z\neq1$. 

In 2009, using Lifshitz's idea of the anisotropic scaling, 
Ho{\v r}ava proposed a renormalizable theory of gravity \cite{Horava:2009uw}, 
which is now called the Ho{\v r}ava-Lifshitz (HL) gravity.
This gravity theory was constructed to be power-counting renormalizable 
by utilizing the Lifshitz scaling with $z=3$ in $(1+3)$-dimensions. 
This remarkable property has triggered extensive studies 
in the application of the HL gravity to the cosmological problems 
(for a review, see \cite{Mukohyama:2010xz} and references therein),
the emergence of the dark matter as an integration constant
~\cite{Mukohyama:2009mz} 
and so on.

While the applications of the HL gravity have been extensively studied, 
other types of field theories with the Lifshitz scaling, 
particularly theories of a Lifshitz scalar, 
have also been investigated to solve some problems of particle physics 
(for a review, see \cite{Alexandre:2011kr}) and 
to understand quantum aspects as a toy model of the HL gravity \cite{Visser:2009fg}.   
In recent years, the quantum properties of Lifshitz scalar theories have been studied \cite{Anselmi:2007ri}. 

In this paper, we study the renormalizability of theories of a Lifshitz scalar, as simple examples of Lifshitz scaling theories.
Although theories with the $z=D$ scaling, such as the HL gravity, are expected to be renormalizable from the viewpoint of power-counting, 
it has not yet been rigorously shown that 
all UV divergences can be canceled by a finite number of counter terms. 
We 
explicitly calculate one-loop contributions in theories of a Lifshitz scalar, to understand the structure of UV divergences in Lifshitz-type theories.
We show that the power counting argument works at one-loop order. 

One of the benefits of the Lifshitz scaling is that 
in the discussion of power-counting, 
higher order interaction terms behave well  
due to the different scaling property from that in the standard field theories with the Lorentz symmetry. 
In particular, in $(1+D)$-dimensional theories with the $z=D$ scaling, the scalar field $\phi$ becomes dimensionless (see eq.(\ref{appendix3})), and 
any interaction term with $2z$ or fewer spatial derivatives is renormalizable in the sense of power-counting (see eq.(\ref{lam2})). 
This means that there can be infinitely many types of UV divergent diagrams,
so that we are forced to deal with an infinite number of counter terms. 
This problem can be avoided by prohibiting potentially dangerous interaction terms 
which cause infinitely many UV divergences. 
One natural way is to impose symmetries 
which restrict the possible forms of interaction terms in the Lagrangian. 
In this paper, we consider a simple model of this type: 
a theory of a single Lifshitz scalar field $\phi$ with the shift symmetry, 
i.e. the invariance under the shift\footnote{
The shift symmetry can also be viewed as a $U(1)$ symmetry if $\phi$ is a periodic scalar field: $\phi \sim \phi + 2 \pi$. 
Such a scalar field naturally appears, for instance, as a Nambu-Goldstone mode of a $U(1)$ symmetry breaking.
}
\begin{eqnarray}
\phi \to \phi+c. \label{SS}
\end{eqnarray}
We show at one-loop order that all UV divergences can be canceled by a finite number of counter terms. 


This paper is organized as follows. 
In Sec.\,\ref{LST}, some basic properties of the Lifshitz scaling are briefly reviewed. 
In Sec.\,\ref{ren}, we discuss the importance of symmetries
which make the number of counter terms finite. 
In Sec.\,\ref{action},  $(1+3)$-dimensional Lifshitz scalar actions with the $z=3$ scaling are discussed 
and interaction terms with/without the shift symmetry are presented. 
In Sec.\,\ref{EC}, we calculate certain types of one-loop diagrams.
After seeing some examples in Sec.\,\ref{ECA} and \ref{ECB},
we prove in Sec.\,\ref{ECC} that, at one-loop order, 
all the UV divergences can be absorbed by shifting bare parameters in the action. 
We also discuss higher loop diagrams from the viewpoint of the superficial degrees of divergence
and see that infinitely many types of UV divergences appear in a generic theory, 
whereas there is no UV divergence for $n$-point diagrams with $n > 6$ in the presence of the shift symmetry. 
In Sec.\,\ref{ECD}, we also discuss some non-symmetric theories
which can be obtained by deforming the symmetric theory. 
Finally, we give the summary and discussion in Sec.~\ref{SecSum}.

\section{power-counting} \label{LST}

In this section, we recapitulate the discussion of power-counting. 
In Sec.~\ref{2A}, we determine the scaling dimensions of space, 
time and the scalar field $\phi$ from the quadratic action. 
In Sec.~\ref{Secit}, from these dimensions, 
we construct all interaction terms which are harmless 
in the sense of non-negative dimensionality of their coupling constants.

\subsection{Lifshitz scaling}\label{2A}

The primary reason of studying theories with the Lifshitz scaling is to
construct theories with an unconventional scaling property 
by modifying the scaling between space and time.
In the relativistic theory, in the natural units $c=\hbar=1$, 
the scaling dimensions of both time and spatial directions are identical, 
i.e. $[dx]=[dt]:=[E]^{-1}$, where $[E]$ is the energy scale.
A standard relativistic action for a scalar field $\phi$ takes the form
\begin{eqnarray}
{\mathcal S} = \int dt \, d^Dx  \,  \left\{ \frac{1}{2} \left( \partial_t \phi\right)^2 - \frac{1}{2} \left(\partial_i \phi\right)^2 + \mathcal{L}_{int} \right\} ,
\end{eqnarray}
where $\partial_i$ is the partial derivative with respect to the \emph{spatial} coordinates $x^i~(i=1,\cdots,D)$.   
The scaling dimension of the scalar field $\phi$ can be read from the kinetic part of the action:  
in terms of the energy scale, it is given by $[\phi]=[E]^{(D-1)/2}$.
Using this scaling property of the coordinates and the scalar field, 
we can derive the scaling dimensions of interaction terms. 
This \emph{power-counting} procedure is also applicable to the case with the Lifshitz scaling, 
where interaction terms scale differently from those in relativistic theories.

The Lifshitz scaling of the space-time coordinates is defined by
\begin{eqnarray}
[dt]=[dx]^z ~ (:=[E]^{-1}), \qquad (\mbox{i.e.}\ [\partial_t]=[\partial_i]^z),
\label{scalest}
\end{eqnarray}
where $z$ is a positive integer. 
Because of the different scalings between time and space, 
theories with $z \neq 1$ do not have the Lorentz symmetry. 
A simplest example with this scaling property is 
the following quadratic action violating the Lorentz symmetry:
\begin{eqnarray}
{\mathcal S}=\int dt \, d^Dx \, \left\{ \frac{1}{2} \left(\partial_t \phi\right)^2 - \frac{1}{2} \phi{(-\Delta})^z \phi \right\},
\label{2nd}
\end{eqnarray}
where $\Delta~(:=\partial_i^2)$ is the spatial Laplacian.\footnote{
The number of time derivatives is kept to second order
to evade the ghost problem associated with higher-order time derivatives.}
From this free action, we can read the scaling dimension of $\phi$:
\begin{eqnarray}
[\phi]=[E]^{(D-z)/2z} 
\label{appendix3}.
\end{eqnarray}
For $z=1$, the action (\ref{2nd}) reduces to that of the free Lorentz-symmetric theory.
On the other hand, in the case of $z=D$, the field $\phi$ becomes dimensionless, 
and hence the scaling dimensions of interaction terms are insensitive to the number of $\phi$. 
Consequently, such theories have the interesting properties that 
any interaction terms with the same number of spatial derivatives have the identical dimension. 
Hereinafter, we investigate the theory with $z=D$ (in UV regime).

%
%
%

\subsection{Power-counting renormalizable interactions}\label{Secit}

Here, we construct interaction terms expected to be harmless from the viewpoint of the power-counting renormalizability. 
From an analogous discussion to that for the Lorentz symmetric theories, 
we expect that the power-counting is helpful to check 
the renormalizability also in the case of Lifshitz scaling theories. 
In the argument of the power-counting,  
we look at the dimensions of coupling constants 
and if all of them are non-negative, such a theory is supposed to be renormalizable.
Similarly, we construct apparently renormalizable interaction terms in the Lifshitz scalar theory. 
In the following, we restrict ourselves only to the case with $z=D$ where the field $\phi$ is dimensionless.

First, let us look at the $n$-point interaction term which takes the form 
\begin{eqnarray}
S_{n}=\lambda_n \int dt \, d^Dx \ \phi^n.
\end{eqnarray}
From the scaling dimensions of the space-time coordinates (\ref{scalest}) 
and the field $\phi$ (\ref{appendix3}), 
we find the dimension of $\lambda_n$ is given by 
\begin{eqnarray}
[\lambda_n]=[E]^{2}. \label{lam}
\end{eqnarray}
Since $[\lambda_n]$ is positive for any $n$, 
any term without derivatives is renormalizable in the sense of power-counting.

Next, let us look for the possibility of the interaction terms with derivative couplings:
\begin{eqnarray}
S_{a,n}=\lambda_{a,n} \int dt \, d^Dx ~ \partial_i^a \phi^n.\label{an}
\end{eqnarray}
Note that this simplified expression denotes  
an $n$-point interaction which contains $a$ spatial derivatives acting on $n$ $\phi$'s.
The dimension of $\lambda_{a,n}$ is calculated to be 
\begin{eqnarray}
[\lambda_{a,n}]=[E]^{(2z-a)/z}.\label{lam2}
\end{eqnarray}
Therefore, any interaction (\ref{an}) with $a \le 2z$ is power-counting renormalizable.

The symbol $z$ is sometimes (especially in cosmology) used to indicate  
both the dynamical exponent and the number of spatial derivatives; 
a term with $a$ spatial derivatives is called a $z=a/2$ term. 
In this paper, to avoid the confusion, 
we use the symbol $z_i$ to indicate the number of derivatives, 
while $z$ is used for the dynamical critical exponent. 


%
%

\section{Renormalizability and importance of symmetry}\label{ren}

We have seen above that in the theory with $z=D$, 
any interaction terms with $2z$ or fewer spatial derivatives but with any $n$ power of $\phi$ are renormalizable 
from the viewpoint of \emph{power-counting}. 
The UV divergence behavior of such terms is demonstrated by an example of 1-loop graph in Sec.~\ref{ECB}.
It means that we need to introduce an \emph{infinite number of counter terms}, 
and hence an infinite number of parameters appear in the theory. 
There appear two possible problems related to these infinitely many counter terms. 
We propose a resolution to these difficulties.

One of the important characteristics of renormalizable theories is \emph{predictability}. 
In the case of finitely many interaction terms, 
all parameters can be fixed by a finite number of renormalization conditions, 
uniquely determining physical quantities, 
i.e. theories with a finite number of parameters have predictability. 
On the other hand, if there are infinitely many parameters, 
it is impossible to fix all of them with a finite number of experimental data. 
Such theories lose complete predictability.\footnote{
It is possible to predict some quantities by fixing a portion of parameters 
as in the case of, for example, the chiral perturbation theory.}


If a theory admits infinitely many power-counting renormalizable terms, 
the RG flow generates all such terms.  
They remain in the effective action in the low energy limit. 
It is not allowed to arbitrarily truncate the action and 
we have to keep track of the RG flows of all the coupling constants 
of renormalizable interactions.
It is practically hard to deal with such a theory 
unless we have some handle on the infinite dimensional RG flow equations.\footnote{
The non-linear sigma model in two dimensions can be viewed as one such example, 
and there are still many discussions (see, for instance, \cite{Ketov}).}

Both of the above concerns come from the fact that 
all power-counting renormalizable terms appear as counter terms 
for divergent loop diagrams. 
We can avoid this problem by removing renormalizable 
but harmful interaction terms which lead to infinitely many divergences.
Restricting the forms of interaction terms by hand may be possible but it is too ad-hoc, 
one natural way is to impose symmetries on the action. 
The HL gravity is one such example: 
the invariance under the foliation-preserving diffeomorphism 
prohibits infinitely many counter terms. 
Another example is the Lifshitz scalar theory with the shift symmetry for $\phi$, see eq.(\ref{SS}), 
which we will focus on in this paper. 
Both problems mentioned above will be  shown to be avoided in this theory.

Considering the coupling with other fields, it is hard to restrict the forms of interaction by hand and symmetries are more helpful. 
Without symmetries, we hardly predict which terms are needed for the renormalization.
The importance of symmetries for the renormalizability
may have important implications for the construction of a consistent model with the Lifshitz scaling.
For instance, when we introduce matter fields in the HL gravity, 
the requirement of symmetries will be helpful as a guide to construct a consistent action.
Without any symmetry, scalar field theories with the $z=D$ scaling 
in general have the problems associated with infinitely many counter terms
since the harmful interaction terms can be generated through the interactions with other fields. 
The symmetry requirement imposes constraints not only on the form of action for matter fields, 
but also on their couplings to the gravity and other fields. 

\section{$z=3$ Lifshitz scalar in $(1+3)$-dimensional space-time}\label{action}

In this section, we construct 
the power-counting renormalizable action of the $z=3$ Lifshitz scalar theory 
in the $(1+3)$-dimensional flat space-time. 
In theories with $z=3$, the action contains couplings with $z_i \leq 3$, 
that is, interaction terms with six or fewer spatial derivatives. 
Since terms with $z_i < 3$ become less important 
than those with $z_i=3$ in the high energy limit,
the quadratic action implies that
that such a theory has the scaling $[dt]=[dx]^3$ in the UV regime.
Then, as we have seen in Sec.\,\ref{Secit}, 
only $z_i \le 3$ terms are renormalizable.
First, we enumerate all the possible third-order terms with $z_i=3$, 
and then we show how the shift symmetry (\ref{SS}) 
reduces the number of allowed terms. 
After that, we explicitly write down all the possible $z_i \le 3$ terms with the shift symmetry.

The most general $z_i \le 3$ quadratic part of the action takes the form
\begin{eqnarray}
S_{2}= \frac{1}{2} \int dt d^3x \Big[ (\partial _t \phi)^2 + \phi {\cal O}\phi \Big], \label{2ndac}
\qquad \mbox{with} \qquad
{\cal O} =\Delta^3 - s \Delta^2 + c_s^2 \Delta - m^2.
\end{eqnarray}
Note that we can always rescale the field $\phi$ and the coordinates to normalize
the coefficients in front of $(\partial_t \phi)^2$ and $\phi \Delta^3 \phi$.
In the UV regime, since $\Delta^3$ term becomes dominant in the differential operator ${\cal O}$, 
we can ignore the other terms and hence $S_2$ reduces to the $z_i=3$ quadratic action (\ref{2nd}).

Recalling that $[\phi]=0$ in the present case ($z=D=3$), 
any interaction term with $z_i \le 3$ satisfies 
the condition of the power-counting renormalizability.
Such terms are written as
\begin{eqnarray}
S_{int}=\sum^\infty_{n=3}{S_{z_i \le 3,n}}\ ,
\end{eqnarray}
where $n$ indicates the order in $\phi$. 
In this action, each term can have independent coupling constant 
and hence an infinite number of independent counter terms would be needed to renormalize the theory. 
 
To ensure the theory to be renormalizable by finitely many counter terms, 
we introduce the shift symmetry (\ref{SS}). 
In a shift symmetric action, the scalar field $\phi$ must always appear 
in combination with space-time derivatives.
Let us see how the shift symmetry restricts the form of the action 
in the simple case of cubic terms with $z_i=3$. 
The most general $z_i=3$ cubic terms can be written as follows
\begin{eqnarray}
S_{3,3} = 
\int dt d^3x \, \left\{ 
\alpha_1 \phi^2 \Delta^3 \phi + 
\alpha_2 (\Delta^2 \phi) (\partial_i \phi)^2 +
\alpha_3 (\Delta \phi)^3\right\}, 
\label{appendixS3} 
\end{eqnarray}
where we have ignored terms related by the integration by parts and 
chosen the above three terms as independent interactions~\cite{Izumi:2010yn}.
In the last two terms, all $\phi$'s have derivatives acting on them
and hence these two terms are invariant under the shift 
\begin{eqnarray}
\phi\to\phi+c.
\end{eqnarray} 
On the other hand, the first term is not invariant, 
that is, the shift symmetry forces $\alpha_1$ to vanish.

The shift symmetric four-point interaction terms with $z_i=3$ 
were derived by Izumi, Kobayashi and Mukohyama~\cite{Izumi:2010yn}:
\begin{eqnarray}
S_{3,4} = \int dt d^3x \, 
\bigl\{
\alpha_4 \left( \Delta\phi \right)^2 \left(\partial_i \phi\right)^2 +
\alpha_5 \left( \partial_i\partial_j\phi \right)^2 \left(\partial_k \phi\right)^2 + 
\alpha_6 (\partial_i \partial_j \partial_k \phi) (\partial_i \phi) (\partial_j \phi) (\partial_k \phi)\bigr\}. \label{4ac}
\end{eqnarray}
There is one shift symmetric term 
in each of $z_i=3$ five and six-point interaction terms: 
\begin{eqnarray}
&&S_{3,5} = \alpha_7 \int dt d^3x \, \left( \partial_i \phi\right)^2\left( \partial_j \phi\right)^2 \Delta \phi, \nonumber
\\
%
%
&&S_{3,6} = \alpha_8 \int dt d^3x \,
\left( \partial_i \phi\right)^2\left( \partial_j \phi\right)^2 \left( \partial_k \phi\right)^2. 
\label{appendix17}
\end{eqnarray}
Higher-order terms never appear since the shift symmetry requires 
that the order of $\phi$ is less than or equal to the number of spatial derivatives, 
which should be less than or equal to $2z=6$
so that the theory is power-counting renormalizable.
For $z_i \le 2$ terms with the shift symmetry, we have only two possibilities:
\begin{eqnarray}
&&S_{2,3} = \alpha_9 \int dt d^3x \, \left( \partial_i \phi\right)^2 \Delta \phi, \nonumber
\\
&&S_{2,4} = \alpha_{10} \int dt d^3x \, \left( \partial_i \phi\right)^2\left( \partial_j \phi\right)^2. \label{z2}
\end{eqnarray}
There is no possible $z_i = 0, 1$ interaction term. 
Therefore, the shift symmetric action for the $z=3$ Lifshitz scalar field consists of thirteen terms: 
(\ref{2ndac}) with $m=0$, (\ref{appendixS3}) with $\alpha_1=0$, (\ref{4ac}), (\ref{appendix17}) and (\ref{z2}).\footnote{
Note that the last term in (\ref{2ndac}) is symmetric 
since the variation under the shift is a total derivative term.
}
Since the number of interaction terms allowed by the shift symmetry is finite, 
the discussion of the power-counting implies that the UV divergences in this theory 
can be absorbed by finitely many counter terms. 

\section{Divergences of one-loop diagrams} \label{EC}

In this section, we investigate the structure of UV divergences in the perturbative expansion. 
Although it is hard to  evaluate exactly one-loop contributions,  
we can easily extract the UV behavior of loop integrals which is needed for the check of renormalizability and counter terms. 
To this end, we concentrate on the high-energy contributions 
and expand the integrands of loop integrals with respect to external momenta. 
Since now we are interested only in the UV-divergences, 
we ignore the contributions from $z_i \le 2$ part in the propagator,  i.e.
we approximate the propagator as
\begin{equation}
D(\omega, \mathbf{p}) = \frac{1}{\omega^2 - |\mathbf{p}|^6}.
\end{equation}

\subsection{Three-point one-loop diagram} \label{ECA}

\begin{figure}[tbp]
  \begin{center}
    \includegraphics[keepaspectratio,height=50mm]{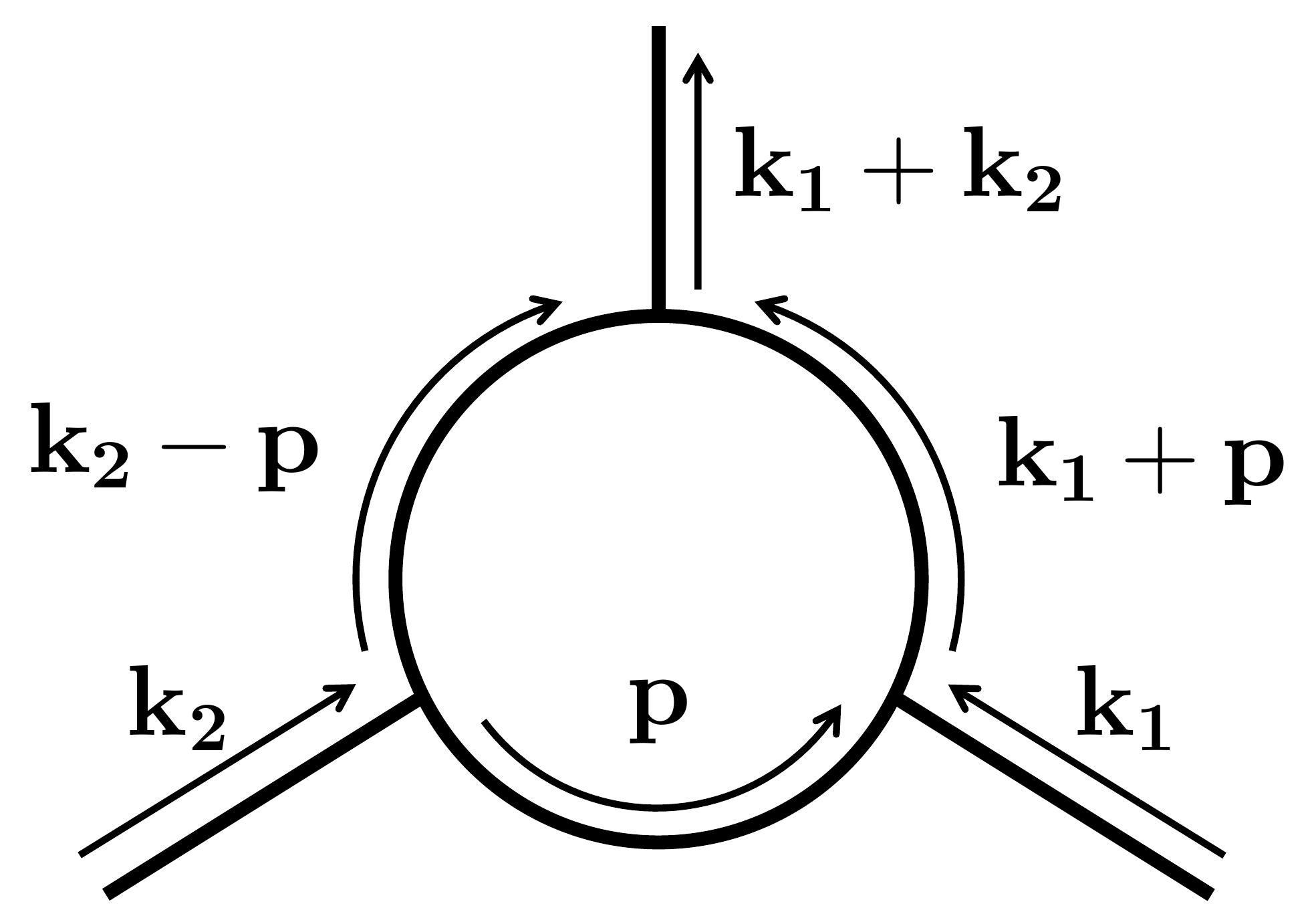}
  \end{center}
  \caption{one-loop diagram of the 3-point function}
  \label{fig:3point.pdf}
\end{figure}

First, to understand the structure of loop integrals, 
we consider a simple example of the one-loop three point diagram (Fig.\,\ref{fig:3point.pdf})
with three vertices corresponding to the following $z_i=3$ cubic interaction term:
\begin{eqnarray}
S_3 =  \int dt d^3x \, \alpha_3  (\Delta \phi)^3 . \label{alpha3}
\end{eqnarray}
After the Wick rotation, the contribution form the one-loop diagram Fig.\,\ref{fig:3point.pdf} becomes 
\begin{eqnarray}
\Gamma_3 \hspace{-2mm} &=& \hspace{-2mm} \int d\omega_p d^3 p \, 
\left(\alpha_3 |\mathbf{k}_1|^2 |\mathbf{p}|^2 |{\bf k_1}+{\bf p}|^2 \right) \frac{1}{\omega_p^2 + |\mathbf{p}|^6} 
\left(\alpha_3 |\mathbf{k}_2|^2 |\mathbf{p}|^2 |{\bf k_2}-{\bf p}|^2\right) \frac{1}{(\omega_2 - \omega_p)^2+|{\bf k_2}-{\bf p}|^6} \nonumber\\
&& \qquad\qquad\times
\left(\alpha_3 |{\bf k_1}+{\bf p}|^2 |{\bf k_2}-{\bf p}|^2 |{\bf k_1}+{\bf k_2}|^2\right) \frac{1}{(\omega_1+\omega_p)^2+|{\bf k_1}+{\bf p}|^6},
\label{Gamma3}
\end{eqnarray}
where we have ignored the symmetry factor since it is not essential in our discussion.
Let us focus on the domain of integration of
 large loop momenta  compared with the external momenta, 
i.e. $|\mathbf{k}_i| \ll |\mathbf{p}|,~\omega_i \ll \omega_p$. 
We expand each factor in the loop integral with respect to $\mathbf{k}_i$ and $\omega_i$. 
At the leading order, the loop integral is given by
\begin{eqnarray}
\Gamma_3 = \alpha_3^3 |\mathbf{k}_1|^2 |\mathbf{k}_2|^2 |\mathbf{k}_1 + \mathbf{k}_2|^2 \int d \omega_p d^3 p \, 
\frac{|\mathbf{p}|^{12}}{(\omega_p^2+|\mathbf{p}|^6)^3} \Bigg[ 1 + 
\mathcal O \left( \frac{|\mathbf{k}_i|}{|\mathbf{p}|},\frac{\omega_i}{\omega_p} \right) \Bigg].
\end{eqnarray}
Let us rewrite this integral in terms of the new variables defined by
\begin{eqnarray}
\omega_p = \rho^3 \cos \theta,  ~~~ |\mathbf{p}|^3 = \rho^3 \sin \theta, ~~~
\hat{\mathbf p} = \frac{\mathbf{p}}{|\mathbf p|}, \hspace{10mm} 
(|\omega_i|^{\frac{1}{z}},\,|\mathbf k_i| \ll \rho \leq \Lambda ,~ 0 \leq \theta \leq \pi),
\label{trans}
\end{eqnarray}
where $\Lambda$ is a momentum cutoff. 
Integrating over the angle $\hat{\mathbf{p}}$, 
we find that the divergent part of this diagram is
\begin{eqnarray}
\Gamma_3 &=& 4\pi \alpha_3^3 |\mathbf{k}_1|^2 |\mathbf{k}_2|^2 |\mathbf{k}_1 + \mathbf{k}_2|^2 
\int \frac{d \theta d \rho}{\rho} \left[ \sin^4 \theta + \mathcal O\left( \frac{|\mathbf{k}_i|^2}{\rho^{2}} \right) \right] \notag \\
&\sim& \frac{3\pi \alpha_3^3}{2} |\mathbf{k}_1|^2 |\mathbf{k}_2|^2 |\mathbf{k}_1 + \mathbf{k}_2|^2 \, \log \Lambda.
\end{eqnarray}
This divergence can be absorbed by the original action (\ref{alpha3}). 
The next leading order contributions, which have $k_i^8$ dependence, are UV-finite 
since the higher order terms are more suppressed in the high energy regime.
We have shown that $z_i\ge 4$ counter terms are not required.

\subsection{$n$-point one-loop diagrams}\label{ECB}

The power-counting argument in Sec.\,\ref{ren} implies  
that without symmetries counter terms of any order in $\phi$ with $z_i \le 3$ appear, 
whereas theories with the shift symmetry do not have $n$-point interaction terms with $n>6$. 
The shift symmetry (\ref{SS}) 
imposes the condition that all $\phi$'s must appear in combination with derivatives in the action. 
On the other hand, each term can have at most six spatial derivatives in the theories with $z=3$.
Therefore, we expect that the difference of UV divergences 
in theories with or without the shift symmetry appears at $n$-point diagrams with $n > 6$. 
In the following, we evaluate $n$-point diagrams with certain types of vertices 
and show the difference between theories with and without the shift symmetry.

\subsubsection{Case without symmetry}
\begin{figure}[tb]
\begin{minipage}{0.495\hsize}
\centering
\includegraphics[keepaspectratio=true,height=50mm]{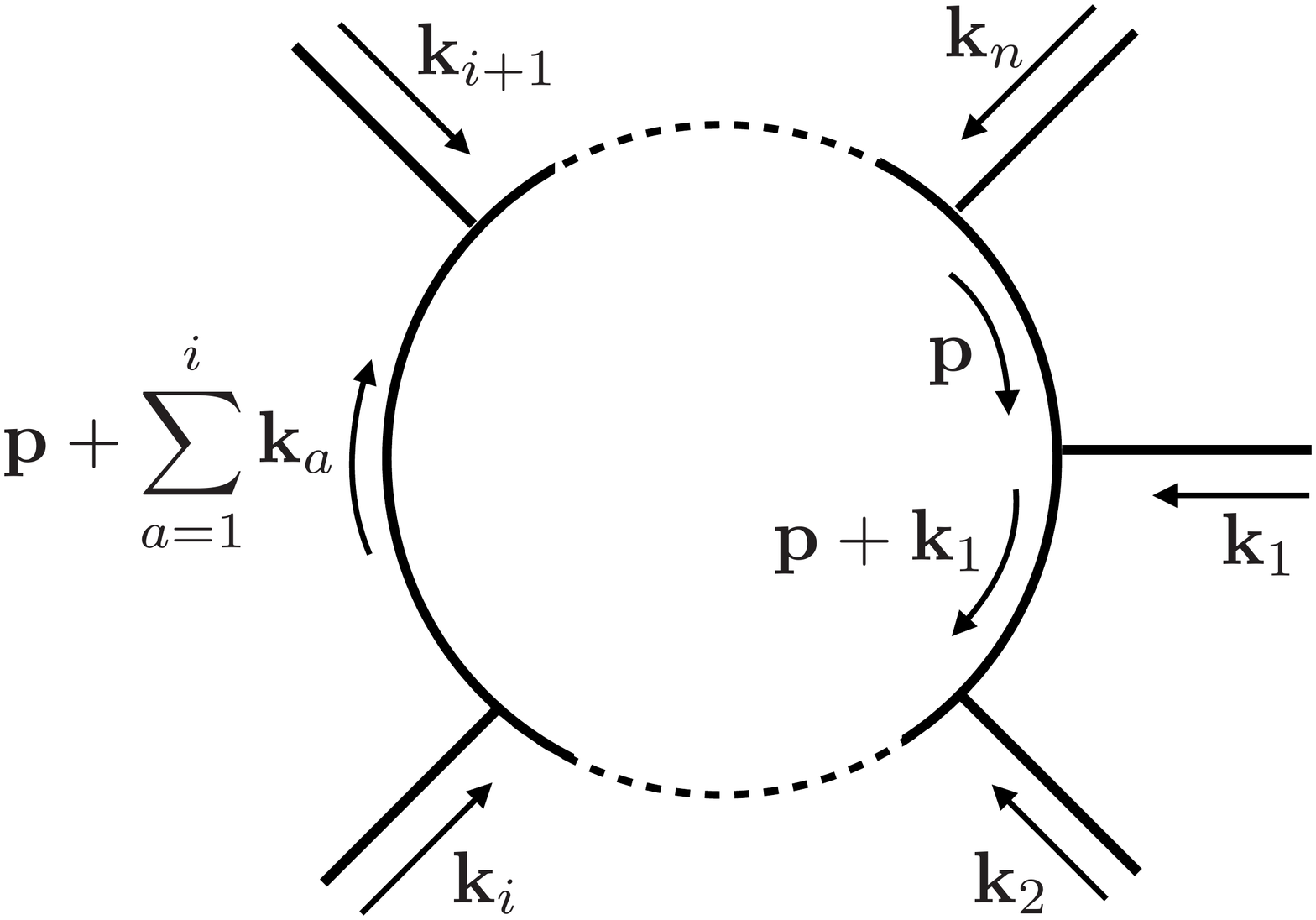}
\caption{$n$-point diagram with 3-point vertices}
\label{fig:npoint1}
\end{minipage}
\begin{minipage}{0.495\hsize}
\centering
\includegraphics[keepaspectratio=true,height=50mm]{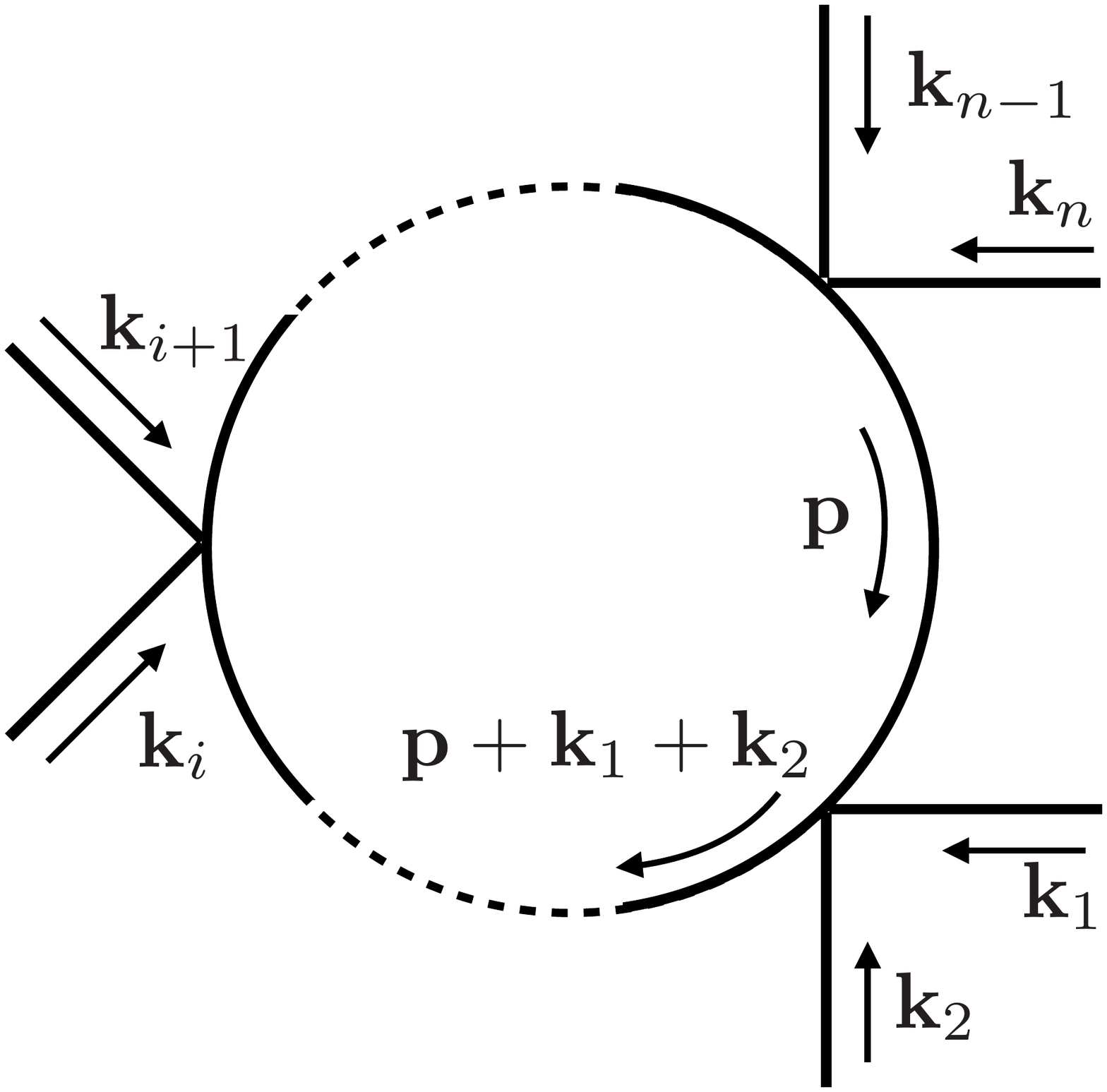}
\caption{$n$-point diagram with 4-point vertices}
\label{fig:npoint2}
\end{minipage}
\end{figure}

First, we consider diagrams in a theory without the shift symmetry.
As an example, let us consider the $n$-point one-loop diagram (Fig.\,\ref{fig:npoint1})
in which all the vertices are those for the following third-point interaction:
\begin{eqnarray}
S_3 = \alpha_1 \int dt d^3 x \ \phi^2 \Delta^3 \phi.
\end{eqnarray}
Since the leading order term is $|\mathbf{p}|^6$ for all the vertices, 
the loop integral can be expanded as
\begin{eqnarray}
\Gamma_n &=& \alpha_1^n \int d \omega_p d^3p \frac{p^{6n}}{(\omega_p^2 + p^6)^n} 
\left[ 1 + O\left( \frac{k}{p} \right) \right] \notag \\
&\propto& \alpha_1^n \int d \rho d \theta \, \rho^5 \left[ \sin^{2n} \theta + O\left( \frac{k}{\rho} \right)\right]
~\sim~ \alpha_1^n \Lambda^6.
\end{eqnarray}
The leading order contribution is of order $\Lambda^6$, 
so that there are UV divergences also in several higher order terms. 
These UV divergences can be canceled by the counter terms of the form
\begin{eqnarray}
S_{conter} \sim \sum_{n=0}^\infty \alpha_1^n \int dt d^3 x \left( \Lambda^6 \phi^n + \mbox{terms with derivatives} \right).
\end{eqnarray}
We have shown that infinitely many counter terms need to be introduced. 

\subsubsection{Case with shift symmetry}

Next, let us consider the case with the shift symmetry. 
Here, we consider the $n$-point diagram ($n=2m$, Fig.\,\ref{fig:npoint2}) which 
consists only of the following shift symmetric four point interaction:
\begin{eqnarray}
S_4 = \alpha_4 \int dt d^D x \, (\Delta \phi)^2 (\partial_i \phi)^2 .
\end{eqnarray} 
In this case, the leading order term of each vertex is proportional to 
$\mathbf{k}_{2i} \cdot \mathbf{k}_{2i-1} |\mathbf{p}|^4$ so that,  
compared with the previous case, 
the integrand has a suppressed behavior for large $|\mathbf{p}|$.
Expanding in terms of the external momenta, we find that 
\begin{eqnarray}
\Gamma_n &=& \alpha_4^m \left( \prod_{i=1}^m \mathbf{k}_{2i} \cdot \mathbf{k}_{2i-1} \right) 
\int d\omega_p d^3 p \frac{p^{4m}}{\left(\omega_p^2+p^6 \right)^m} \left[  1 + O\left( \frac{k}{p} \right) \right] \nonumber\\
&\propto& \alpha_4^m \left( \prod_{i=1}^m \mathbf{k}_{2i} \cdot \mathbf{k}_{2i-1} \right)  
\int d \rho d \theta \rho^{5-2m} \left[ \sin^{\frac{2m}{3}} \theta +O\left( \frac{k}{\rho} \right) \right].
\end{eqnarray}
The diagram becomes less divergent for larger $m$ and there is no divergence for $n=2m>6$. 
Therefore, the UV divergence from this type of diagrams can be canceled by a finite number of counter terms.

\subsection{Finite number of counter terms in theories with shift symmetry} \label{ECC}

In the examples above, we have seen that 
without the shift symmetry there are infinitely many UV divergences
whereas there is no divergence in the $n$-point diagram with $n>6$ in the theory with the shift symmetry. 
In this section, we show in general that, at one-loop order, 
only $z_i \le 3$ counter terms are required and 
there are finitely many divergences in the shift symmetric case.

\subsubsection{general discussion for one-loop diagrams}
In the theory with $z=3$ scaling the action consists of $z_i \le 3$ interaction terms, 
which have at most six spatial derivatives. 
Since the numbers of internal lines and vertices are always the same in any one-loop diagram,  
it has the following schematic form
\begin{eqnarray}
\Gamma = \int d\omega_p d^3 p \, V_1 D_1 V_2 D_2 \cdots V_q D_q.
\end{eqnarray} 
Here, all the vertices $V_I~(I=1,2\cdots,q)$ are polynomials of the momenta 
of order less than or equal to six: 
\begin{eqnarray}
V_I = \sum_{\{a_i\}} C_{\{a_i\}} \, p^{a_0} k_1^{a_1} k_2^{a_2} \cdots k_n^{a_n} \qquad
\mbox{with} \qquad
\sum_{i=0}^{n} a_i \le 6 \qquad
\mbox{and} \qquad
a_i \ge 0,\label{ver}
\end{eqnarray}
where the coefficients $C_{\{a_i\}}$ are functions of coupling constants and 
relative angles of momenta such as ${\bf \hat p \cdot \hat k_i}$ and ${\bf \hat k_i \cdot \hat k_j}$. 
The propagators for internal lines can be expanded as 
\begin{eqnarray}
D_I ~=~ \frac{1}{\omega_p^2 + p ^6 + \cdots} ~=~ \frac{1}{\omega_p^2 + p ^6}
\sum C_{\{b, c_i, d_i \}} \frac{\omega_p^{c_0} \, p^{d_0}}{(\omega_p^2 + p ^6)^b} 
\prod_{i=1}^n \omega_i^{c_i} k_i^{d_i} , \label{pro}
\end{eqnarray}
where the coefficients $C_{\{b, c_i, d_i \}}$ are functions of the relative angles and parameters of the model. 
The summation in eq.(\ref{pro}) is taken over all non-negative integers $b$, $c_i$ and $d_i$ satisfying  
\begin{eqnarray}
6b \ge \sum_{i=0}^n ( 3 c_i  + d_i ).
\label{bc}
\end{eqnarray}
Then, combining (\ref{ver}) and (\ref{pro}), 
we find that any one-loop contribution takes the form
\begin{eqnarray}
\Gamma &=& \int d\omega_p d^3 p \sum \tilde C_{\{\tilde b, \tilde c_i,\tilde d_i\}} 
\frac{\omega_p^{\tilde c_0} \, p^{\tilde d_0}}{(\omega_p^2 + p ^6)^{\tilde b}} 
\prod_{i=1}^n \omega_i^{\tilde c_i}  k_i^{\tilde d_i}, \label{gam}
\end{eqnarray}
where the summation is taken over non-negative integers $\tilde b$, $\tilde c_i$ and $\tilde d_i$ satisfying
\begin{eqnarray}
6 \tilde b \ge \sum_{i=0}^n (3 \tilde c_i  + \tilde d_i), 
\label{eq:cond}
\end{eqnarray}
With the same change of variables as (\ref{trans}), 
eq.(\ref{gam}) can be rewritten as
\begin{eqnarray}
\Gamma = \int d \rho d \theta d^2 \hat p \, \sum \tilde C_{\{\tilde b,\tilde c_i,\tilde d_i\}} \,
\rho^{5 - 6 \tilde b + 3 \tilde c_0 + \tilde d_0} (\cos\theta)^{\tilde c_0} (\sin\theta)^{\frac{\tilde d_0}{3}} 
\prod_{i=1}^n \omega_i^{\tilde c_i} k_i^{\tilde d_i}.
\label{eq:npoint}
\end{eqnarray}
Therefore, the terms with $5 - 6 \tilde b + 3 \tilde c_0 + \tilde d_0 \geq -1$ have UV divergences.
Note that the integration with respect to $\hat p$ and $\theta$ never gives divergence. 
Taking into account \eqref{eq:cond}, we find that UV divergences appear in the terms with
\begin{eqnarray}
\sum_{i=1}^n (3 \tilde c_i  + \tilde d_i) \leq 6.
\label{eq:res}
\end{eqnarray}
This inequality shows that we do not need to introduce $z_i > 3$ terms to absorb the divergences, 
since $\sum_i  \tilde c_i$ and $\sum_i \tilde d_i$ are the numbers of time and spatial derivatives in the counter terms.  
The above discussion also implies that any $n$-point diagram can have UV divergence,
irrespective of the number of external lines. 
Therefore, in general, we are forced to deal with infinitely many counter terms 
to absorb all types of UV divergences. 

On the other hand, in the theory with the shift symmetry, 
at least one spatial derivative is associated with each external line.  
Consequently, for any $n$-point diagrams, 
the numbers of the external momenta in \eqref{eq:npoint} are all non-zero,
that is, 
\begin{eqnarray}
\tilde d_i \geq 1 ~~~\Longrightarrow~~~ \sum_{i=1}^n \tilde d_i \geq n. 
\label{eq:external}
\end{eqnarray} 
This implies that the condition \eqref{eq:res} cannot be satisfied for $n > 6$,  
so that there is no UV divergence in $n$-point diagrams with $n > 6$.
Therefore, at one-loop order, 
all the UV divergences can be canceled by a finite number of counter terms in the shift symmetric case.

\subsubsection{Higher loop diagrams} 
Next, let us discuss the case of higher loop diagrams. 
Since it is not easy to explicitly evaluate higher loop diagrams, 
we restrict ourselves to the discussion of the superficial degrees of divergence.

In a generic Lifshitz scalar theory, a naive estimation shows that 
a diagram has a UV divergence if the following superficial degree of divergence is non-negative: 
\begin{eqnarray}
D_{\rm div} = 2 z (L - P) + \sum_{z_i=0}^{z} 2 z_i V_{z_i}, 
\end{eqnarray}
where $L$ is the number of loop integrals, $P$ is the number of propagators and 
$V_{z_i}~(z_i=0,\cdots,z)$ are the numbers of vertices with $2z_i$ spatial derivatives.  
Note that a generic vertex with $2z_i$ spatial derivatives has $2z_i$ loop momenta, 
so that its contribution to $D_{\rm div}$ is $2z_i$. 
Counting the number of undetermined momenta in the diagram, 
we find that the numbers of loop integrals, propagators and vertices are related as
\begin{eqnarray}
L = P - \sum_{z_i=0}^{z} V_{z_i} + 1. 
\end{eqnarray}
Then, eliminating $L$ and $P$, we can rewrite $D_{\rm div}$ as
\begin{eqnarray}
D_{\rm div} = 2 z - \sum_{z_i=0}^{z} 2(z-z_i) V_{2z_i} .
\end{eqnarray}
This shows that vertices with $z_i = z$ do not change the superficial degree of divergence. 
This is the reason why there are infinitely many types of UV divergences.    

On the other hand, in the case with the shift symmetry, 
we also have to take into account the number of external lines. 
Since each external line attached to a shift symmetric vertex 
reduces $D_{\rm div}$ at least by one, 
the superficial degrees of divergence of 
$n$-point diagrams satisfy the following inequality:
\begin{eqnarray}
D_{\rm div} ~\leq~ 2 z - \sum_{z_i=0}^{z} 2(z-z_i) V_{2z_i} - n ~\leq~ 2 z - n.
\end{eqnarray}
Since $n$-point diagrams with $n>2z=6$ always have negative $D_{\rm div}$, 
they are expected to be finite and hence finitely many counter terms would be sufficient 
for the renormalization at any loop order.

\subsection{Deformations of symmetric theories} \label{ECD}
We have seen in the argument of the superficial degree of divergence that, 
in general, marginal couplings, i.e. interactions with $2z$ spatial derivatives,
give rise to infinitely many divergences. 
By removing such potentially dangerous terms from the action, 
we can construct ``healthy" theories which can be renormalized by a finite number of counter terms.
As we have seen in the above, we can naturally 
restrict the forms of interaction terms by imposing the shift symmetry.  
Here, we discuss the possibility of non-symmetric theories 
which are free from the potentially dangerous terms leading to infinitely many divergences.

One simple example can be obtained by deforming a symmetric theory by new interaction terms 
which can be obtained from shift symmetric terms by replacing spatial derivatives into constants.   
For example, let us consider the deformed interaction obtained 
by replacing the cubic interaction term as
\begin{eqnarray}
(\Delta \phi)^3 \rightarrow (\Delta \phi)^3 + M_2^2 \phi (\Delta \phi)^2 + M_1^4 \phi^2 (\Delta \phi) + M_0^6 \phi^3. 
\end{eqnarray}
This corresponds to the following replacement of the vertex 
\begin{eqnarray}
|\mathbf p_1|^2 |\mathbf p_2|^2 |\mathbf p_3|^2 &\rightarrow&
|\mathbf p_1|^2 |\mathbf p_2|^2 |\mathbf p_3|^2 + 
M_2^2 ( |\mathbf p_1|^2 |\mathbf p_2|^2 + |\mathbf p_2|^2 |\mathbf p_3|^2 + |\mathbf p_3|^2 |\mathbf p_1|^2) \notag \\
&+& M_1^4 ( |\mathbf p_1|^2  + |\mathbf p_2|^2 + |\mathbf p_3|^2 ) + M_0^6.  
\end{eqnarray}
Obviously, the behavior of this deformed vertex in the large momentum region 
is identical to that of original one. 
Furthermore, we can show that if $n$ external lines are attached to this vertex, 
the superficial degree of divergence is reduced at least by $n$ 
as in the case of the original vertex with the shift symmetry. 
Therefore, at one-loop order, $n$-point diagrams with $n>6$ are finite.

The explicit form of the interaction terms of this example is written in 
\begin{eqnarray}
V = V_{z_i=3}^{sym} + m^2_2 f_2 V_{z_i=2}^{sym} + m^4_1 f_4 V_{z_i=1}^{sym} + m^6_0 f_6,
\label{nosym}
\end{eqnarray}
where $f_n$ is an $n$-th order polynomial of $\phi$ and 
$V_{z_i=n}^{sym}$ is a $z_i=n$ interaction term with the shift symmetry. 
We can easily find that these terms can be obtained by the procedure discussed above. 
The shift symmetry is recovered, for instance, if we replace $m^2_2 f_2$ as 
\begin{eqnarray}
m^2_2 f_2 =m^2_2 (\phi^2 + C_1 \phi + C_0) \qquad \to \qquad 
(\partial_i \phi)^2 + \Delta \phi +C_0.
\end{eqnarray} 
Conversely,  $m^2_2 f_2V_{z_i=2}^{sym}$ can be constructed 
from the $z_i\le 3$ interaction terms with the shift symmetry by the  replacement of derivatives.
Therefore, if higher loop corrections do not require harmful counter terms, 
the $z=3$ Lifshitz scalar theory with the potential (\ref{nosym}) can be
renormalized by a finite number of counter terms. 

\section{Summary}\label{SecSum}

In this paper, we have discussed the renormalizability in the Lifshitz scalar field theories. 
In Sec.\,\ref{LST}, we have reviewed theories with Lifshitz scaling 
and the power-counting procedure. 
If the dynamical critical exponent has the same value 
as the number of spatial dimensions (i.e. $z=D$), 
the scalar field becomes dimensionless. 
In this special case, the dimensions of interaction terms 
do not depend on the order of the fields, 
and thus any interaction term with $2z$ or fewer derivatives 
is renormalizable in the sense of the power-counting. 
We have explicitly checked in Sec.\,\ref{EC} that, at one-loop order, 
the power-counting argument for the renormalizability 
gives the correct answer also in the Lifshitz scalar theory. 
To prove it at any loop order, 
we should perhaps follow a similar procedure 
to the Bogoliubov-Parasiuk-Hepp-Zimmermann renormalization scheme. 

The discussion of power-counting shows that, 
in general, theories of a Lifshitz scalar with the $z=D$ scaling, 
an infinite number of counter terms are required.
In Sec.\,\ref{ren}, we have discussed 
the possible problems originating from the infinitely many counter terms.
Because of the necessity of an infinite number of renormalization conditions, 
these types of theories might not have complete predictability. 
Even if the theories with an infinite number of counter terms are well-posed, 
there may be practical problems; 
we need to handle all the infinitely many interactions, 
since all renormalizable terms have to be equivalently taken into account.  
The above possible problems can be avoided 
if only finitely many counter terms are required. 
In Sec.\,\ref{ECC}, we have explicitly shown that, 
at one-loop order, the number of required counter terms is finite 
in the presence of the shift symmetry which prohibits the potentially dangerous interaction terms. 
We can find more candidates of consistent non-symmetric theories, 
if we allow ourselves to discard some power-counting renormalizable but harmful terms by hand. 
However, if such a theory is coupled to other sectors such as the gravity, 
there is no guarantee that the consistency is maintained 
since the dangerous terms which are not protected by symmetries 
can be generated through interactions with other sectors. 

The symmetry requirements may have important implications 
for the construction of a UV-complete theory with the Lifshitz scaling, 
such as the HL gravity with matter fields.
The power-counting arguments allow an infinite number of coupling terms among fields and gravity~\cite{Carloni:2010ji}. 
If symmetries are really essential in the Lifshitz-type theories,  
we can strongly constrain not only the forms of the self-interaction of scalar fields 
but also the coupling terms with gravity. 
We will discuss whether symmetries are essential for the consistency 
by investigating the (non)unitarity of field theories with the Lifshitz scaling~\cite{future1}.

\section*{Acknowledgement}

This work is supported in part by scientific grants from the Ministry of Education under Grant Nos. 24540285 (T.~Inami). T.I. wishes to thank CTS of NTU for partial support. 
K.I. is supported by Taiwan National Science Council under Project No. NSC101-2811-M-002-103.
T. K. benefitted much from his visits to NTU and NTNU. He wishes to thank Pei-Ming Ho for the kind support for the visit, Hsien-chung Kao for the kind acceptance of him. 
He also thanks Kei-ich Maeda for useful discussions.


\begin{thebibliography}{100}



\bibitem{Lifshitz scalar}
E. Lifshitz, On the theory of Second-Order Phase transitions I II Zh. Eksp.Teor. Fiz. 11 (1941) 255, 269.

\bibitem{Horava:2009uw} 
  P.~Horava,
  Phys.\ Rev.\ D {\bf 79}, 084008 (2009)
  [arXiv:0901.3775 [hep-th]].


  P.~Horava,
  JHEP {\bf 0903}, 020 (2009)
  [arXiv:0812.4287 [hep-th]].

\bibitem{Mukohyama:2010xz} 
  S.~Mukohyama,
  Class.\ Quant.\ Grav.\  {\bf 27}, 223101 (2010)
  [arXiv:1007.5199 [hep-th]].
  
  
\bibitem{Mukohyama:2009mz} 
  S.~Mukohyama,
  Phys.\ Rev.\ D {\bf 80}, 064005 (2009)  [arXiv:0905.3563 [hep-th]].  


  K.~Izumi and S.~Mukohyama,
  Phys.\ Rev.\ D {\bf 81}, 044008 (2010)  [arXiv:0911.1814 [hep-th]].  

  K.~Izumi and S.~Mukohyama,
  Phys.\ Rev.\ D {\bf 84}, 064025 (2011)  [arXiv:1105.0246 [hep-th]].  

  
  

  
\bibitem{Alexandre:2011kr} 
  J.~Alexandre,
  Int.\ J.\ Mod.\ Phys.\ A {\bf 26}, 4523 (2011)
  [arXiv:1109.5629 [hep-ph]].
  
 
\bibitem{Visser:2009fg} 
  M.~Visser,
  Phys.\ Rev.\ D {\bf 80}, 025011 (2009)
  [arXiv:0902.0590 [hep-th]].
  
  M.~Visser,
  arXiv:0912.4757 [hep-th].

  M.~Colombo, A.~E.~Gumrukcuoglu and T.~P.~Sotiriou,
  arXiv:1410.6360 [hep-th].
 
\bibitem{Anselmi:2007ri} 
  D.~Anselmi and M.~Halat,
  Phys.\ Rev.\ D {\bf 76}, 125011 (2007)
  [arXiv:0707.2480 [hep-th]].
  
  D.~Anselmi,
  JHEP {\bf 0802}, 051 (2008)
  [arXiv:0801.1216 [hep-th]].
  
  D.~Anselmi,
  Annals Phys.\  {\bf 324}, 874 (2009)
  [arXiv:0808.3470 [hep-th]].
  
  D.~Anselmi,
  Annals Phys.\  {\bf 324}, 1058 (2009)
  [arXiv:0808.3474 [hep-th]].
   
  R.~Iengo, J.~G.~Russo and M.~Serone,
  JHEP {\bf 0911}, 020 (2009)
  [arXiv:0906.3477 [hep-th]].
    
  K.~Farakos and D.~Metaxas,
  Phys.\ Lett.\ B {\bf 707}, 562 (2012)
  [arXiv:1109.0421 [hep-th]].

  D.~L.~Lopez Nacir, F.~D.~Mazzitelli and L.~G.~Trombetta,
  Phys.\ Rev.\ D {\bf 85}, 024051 (2012)
  [arXiv:1111.1662 [hep-th]].
  
  M.~Baggio, J.~de Boer and K.~Holsheimer,
  JHEP {\bf 1207}, 099 (2012)
  [arXiv:1112.6416 [hep-th]].

  M.~Eune, W.~Kim and E.~J.~Son,
  Phys.\ Lett.\ B {\bf 703}, 100 (2011)
  [arXiv:1105.5194 [hep-th]].
  


  
\bibitem{Ketov}
S.~V.~Ketov,
``Quantum Non-linear Sigma-models: From Quantum Field Theory to Supersymmetry, Conformal Field Theory, Black Holes and Strings, "
Springer Berlin Heidelberg.

J.~Zinn-Justin,
``Quantum Field theory and Critical Phenomena, "
Oxford University Press .

  
\bibitem{Izumi:2010yn} 
  K.~Izumi, T.~Kobayashi and S.~Mukohyama,
  JCAP {\bf 1010}, 031 (2010)
  [arXiv:1008.1406 [hep-th]].

\bibitem{Carloni:2010ji} 
  S.~Carloni, E.~Elizalde and P.~J.~Silva,
  Class.\ Quant.\ Grav.\  {\bf 28}, 195002 (2011)  [arXiv:1009.5319 [hep-th]].  

\bibitem{future1}
T.~Fujimori, T.~Inami, K.~Izumi and T.~Kitamura,
in preparation.



\end{thebibliography}
\end{document}